\documentclass[aps,graphicx,prb,twocolumn,groupedaddress]{revtex4}
\usepackage{amssymb}
\usepackage{epsfig}
\usepackage[centertags]{amsmath}
\usepackage{graphicx,subfigure}
\usepackage{latexsym}

\newcommand{\ket}[1]{\left|#1\right>}
\newcommand{\bra}[1]{\left<#1\right|}
\newcommand{\bok}[3]{\left<#1\right|#2\left|#3\right>}

\begin{document}
\title{Magnetic Excitations in La$_2$CuO$_4$ probed by Indirect Resonant Inelastic X-ray Scattering}
\author{Filomena Forte$^{1,2}$, Luuk J. P. Ament$^1$ and Jeroen van den Brink$^{1,3}$}
\affiliation{$^1$ Institute-Lorentz for Theoretical Physics,  Universiteit  Leiden,\\
P.O. Box 9506, 2300 RA Leiden,The Netherlands\\
$^2$ Dipartimento di Fisica ``E. R. Caianiello'', Universit{\`a} di Salerno, I-84081 Baronissi, Salerno, Italy and Laboratorio Regionale SuperMat, INFM-CNR, Baronissi (SA), Italy\\
$^3$ Institute for Molecules and Materials, Radboud Universiteit Nijmegen,\\
P.O. Box 9010, 6500 GL Nijmegen, The Netherlands}
\date{\today}
\pacs{PACS numbers: 78.70.-g 74.72.-h 78.70.Ck 71.27.+a}

\begin{abstract}
Recent experiments on La$_2$CuO$_4$ suggest that indirect resonant inelastic X-ray scattering (RIXS) might provide a probe for transversal spin dynamics. We present in detail a systematic expansion of the relevant magnetic RIXS cross section by using the ultrashort core-hole lifetime (UCL)  approximation. We compute the scattering intensity and its momentum dependence in leading order of the UCL expansion. The scattering is due to two-magnon processes and is calculated within a linear spin-wave expansion of the Heisenberg spin model for this compound, including longer range and cyclic spin interactions. We observe that the latter terms in the Hamiltonian enhance the first moment of the spectrum if they strengthen the antiferromagnetic ordering. The theoretical spectra agree very well with experimental data, including the observation that scattering intensity vanishes for the transferred momenta ${\bf q} = (0,0)$ and ${\bf q} = (\pi,\pi)$. We show that at finite temperature there is an additional single-magnon contribution to the scattering with a spectral weight proportional to $T^3$. We also compute the leading corrections to the UCL approximation and find them to be small, putting the UCL results on a solid basis.  All this univocally points to the conclusion that the observed low temperature RIXS intensity in  La$_2$CuO$_4$ is due to two-magnon scattering.
\end{abstract}

\maketitle

\begin{section}{Introduction}
Indirect Resonant Inelastic X-ray Scattering (RIXS) is rapidly establishing itself as a new probe of electronic excitations in solids. The recent increase in brilliance of synchrotron radiation has made it possible to observe second order scattering processes as indirect RIXS~\cite{Schuelke07,Kotani01,Hasan00,Kim02,Hill98,Isaacs96,Kao96,Inami03,Abbamonte99,Tsutsui03,Doering04,Suga05,Nomura05,Wakimoto05,Collart06,Seo06}. Moreover, the improvements in the instrumental resolution (100 meV is achieved) allow for lower energy scales to be detected, making this technique in principle a powerful instrument to probe the low-lying elementary excitations of solids, for instance magnons~\cite{Hill_tbp,Brink05b}.

In indirect RIXS, the energy of the incoming photons is tuned to match a resonant edge of an atomic transition in the particular system that one sets out to investigate. This resonance corresponds to exciting a core electron to an outer shell. The $K$-edge of transition metal ions is particularly useful since it promotes a 1$s$ core electron to an outer 4$p$ shell, which is well above the Fermi level, so that the X-rays do not cause direct transitions of the 1$s$ electron into the lowest 3$d$-like conduction bands~\cite{Kotani01,Hasan00,Kim02,Hill98,Isaacs96,Kao96,Inami03,Abbamonte99,Tsutsui03,Doering04,Suga05,Nomura05,Wakimoto05,Collart06,Seo06}.

Due to the large energy involved ($\sim$5-10 keV), the core-hole is ultrashortlived and it induces an almost delta function-like potential (in time) on the valence electrons~\cite{Brink06,Brink05a,Ament07}. Consequently, elementary excitations of the valence electrons will screen the local potential, but have litlle time to do so. When the core-hole decays, the system can be left behind in an excited state. By observing the energy and momentum of the outgoing photon, one probes the elementary excitations of the valence electrons including, in particular, their momentum dependence.

In the last few years, considerable theoretical progress has been made to comprehend RIXS spectra~\cite{Isaacs96,Abbamonte99,Tsutsui03,Doering04} and particularly in the understanding of the correlation functions that are measured by indirect RIXS~\cite{Brink05b,Brink05a,Brink06,Ament07}. It is by now well established that indirect RIXS detects the momentum dependence of charge excitations that are related to the electrons and holes in the $d$-shell in for instance the cuprates and manganites. Treating the scattering problem taking the ultrashort core-hole lifetime (UCL) into account has proved that the indirect RIXS intensity is proportional to the dielectric loss function and longitudinal spin excitations of the electrons that couple to the core-hole~\cite{Brink05a,Ament07}.

Recently, RIXS measurements performed by J.P. Hill and coworkers on the high-T$_c$ cuprate superconductor La$_{2-x}$Sr$_x$CuO$_4$ revealed that RIXS is potentially able to detect {\it transversal} spin excitations --magnons~\cite{Hill_tbp}. The experiments show that the magnetic RIXS signal is strongest in the undoped cuprate La$_2$CuO$_4$. The
magnetic loss features are at energies well below the charge gap of this magnetic insulator, at energies where the charge response function $S({\bf q},\omega)$ vanishes, as well as the {\it longitudinal} spin one --which is in fact a higher order charge response function. The proposed scattering mechanism is a two-magnon scattering process in which two spin waves are created~\cite{Hill_tbp,Brink05b}.

In a previous theoretical analysis we have shown that the magnetic correlation function that is measured by indirect RIXS is a four-spin correlation one, probing two-magnon excitations\cite{Brink05b}. This makes indirect RIXS a technique that is essentially complementary to magnetic neutron scattering, which probes single magnon properties and two-spin correlations. In this paper, we present the theoretical framework of Ref.~\onlinecite{Brink05b} in more detail and use it for an analysis of the experimental magnetic RIXS data on perovskite CuO$_2$ layers of La$_2$CuO$_4$.

We expand upon the previous considerations by providing a detailed comparison between the theory and experiment, including also longer range magnetic exchange interactions in the theory --with values known from neutron scattering data. We develop the theory to account also for the effects of finite temperature, which give rise to a non-trivial single-magnon contribution to the RIXS signal. We also compare with the results of Nagao and Igarashi~\cite{Nagao07}, who recently computed the magnetic RIXS spectra based on the theoretical framework of Ref.~\onlinecite{Brink05b}, taking also some of the magnon-magnon interactions into account.

The theory is developed on basis of the ultrashort core-hole lifetime (UCL) expansion. We compute leading order corrections to this expansion and show that they are small. This makes sure that the UCL approximation provides a reliable route to analyze the indirect RIXS spectra.

This paper is organized as follows: in section \ref{sec:general} we obtain an expression for the cross section of the 2D $S=1/2$ Heisenberg antiferromagnet in linear spinwave theory in terms of magnon creation and annihilation operators. In section \ref{sec:T=0} we evaluate the cross section at $T=0$. Section \ref{sec:T>0} concerns the low temperature case. Next, the leading correction to the cross section in the UCL approximation is calculated. Section \ref{sec:conc} is devoted to the concluding remarks.
\end{section}

\begin{section}{Cross Section for Indirect RIXS on a Heisenberg AFM \label{sec:general}}
Recently, J.P. Hill {\it et al.} \cite{Hill_tbp} observed that RIXS on the high $T_c$ superconductor La$_{2-x}$Sr$_x$CuO$_4$ picks up transversal spin dynamics --magnons. In the undoped regime, the RIXS intensity turns out to be highest. The same feature was observed in the related compound Nd$_2$CuO$_4$. These cuprates consist of perovskite CuO$_2$ layers with a hole in the Cu 3$d$ subshell. The low energy spin dynamics of these systems are properly described by a single band Hubbard model at half filling. The strong interactions between holes in the Cu 3$d$ subshells drive these materials into the Mott insulating regime, where the low energy excitations are the ones of the $S=1/2$ 2D Heisenberg antiferromagnet:
\begin{equation}
    H_0 = \sum_{i,j} J_{ij} {\bf S}_i \! \cdot \! {\bf S}_j
\end{equation}
with $J_{ij} \approx 146 $ meV for nearest neighbors~\cite{Coldea01}. The superexchange integral $J_{ij}$ is determined from the virtual hopping processes concerning sites $i$ and $j$: $J_{ij} = 4 t^2_{ij}/U$. Here $t_{ij}$ is the hopping amplitude and $U$ is the Coulomb repulsion between two 3$d$ electrons on the same site. In the antiferromagnetic groundstate, the Hamiltonian can be bosonized in linear spinwave theory (LSWT) where $S^z_i \mapsto 1/2 - a^{\dag}_i a_i,\; S^+_i \mapsto a_i$ and $S^-_i \mapsto a^{\dag}_i$ for $i \in A$ ($A$ being the sublattice with spin-up) and $S^z_j \mapsto b^{\dag}_j b_j - 1/2,\; S^+_j \mapsto b^{\dag}_j$ and $S^-_j \mapsto b_j$ for $j \in B$ (the spin-down sublattice). A Bogoliubov transformation in reciprocal space is necessary to diagonalize $H_0$:
\begin{align}
    \alpha_{\bf k} &= u_{\bf k} a_{\bf k} + v_{\bf k} b^{\dag}_{-{\bf k}}, \label{eq:alpha} \\
    \beta_{\bf k} &= u_{\bf k} b_{\bf k} + v_{\bf k} a^{\dag}_{-{\bf k}} \label{eq:beta}
\end{align}
with
\begin{equation}
    u_{\bf k} = \sqrt{\frac{J^{AB}_{\bf 0}-J^{AA}_{\bf 0}+J^{AA}_{\bf k}}{2 \sqrt{\left( J^{AB}_{\bf 0}-J^{AA}_{\bf 0}+J^{AA}_{\bf k} \right)^2 - \left(J^{AB}_{\bf k}\right)^2}}+\frac{1}{2}}
\end{equation}
and
\begin{equation}
    v_{\bf k} = \text{sign} (J^{AB}_{\bf k}) \sqrt{u^2_{\bf k}-1}
\end{equation}
where $J^{XY}_{\bf k}$ is the Fourier transform of those terms in $J_{ij}$ connecting a site in sublattice $X$ to a site in $Y$. For interactions up to third nearest neighbors we get
\begin{align}
    J^{AB}_{\bf k} = &J \left( \cos a k_x + \cos a k_y \right) \\
    J^{AA}_{\bf k} = J^{BB}_{\bf k} = &2J' \cos a k_x \cos a k_y + \nonumber \\
    &J'' \left( \cos 2 a k_x + \cos 2 a k_y \right)
\end{align}
with $a$ the lattice constant and $J, J', J''$ the first through third nearest neighbor couplings. The final linear spinwave Hamiltonian in terms of boson operators is
\begin{equation}
    H_0 = \text{const}+\sum_{\bf k} \epsilon_{\bf k} \left( \alpha^{\dag}_{\bf k} \alpha_{\bf k} + \beta^{\dag}_{\bf k} \beta_{\bf k} \right)
\end{equation}
with $\epsilon_{\bf k} = \sqrt{(J^{AB}_{\bf 0}-J^{AA}_{\bf
0}+J^{AA}_{\bf k})^2-\left(J^{AB}_{\bf k}\right)^2}$.

Our aim is to understand how this picture changes when doing indirect RIXS. In RIXS, one uses X-rays to promote a Cu 1$s$ electron to a 4$p$ state. For an ultrashort time, one creates a core-hole at a certain site which lowers the Coulomb repulsion $U$ on that site with an amount $U_c$. We assume that the core-hole potential is local, i.e. it acts only at the core-hole site. This approximation is reasonable as the Coulomb potential is certainly largest on the atom where the core-hole is located. Moreover, we can consider the potential generated by both the localized core-hole and photo-excited electron at the same time. As this exciton is a neutral object, its monopole contribution to the potential vanishes for distances larger than the exciton radius. The multi-polar contributions that we are left with in this case are generally small and drop off quickly with distance.

The strong core-hole potential in the intermediate state alters the superexchange processes between the 3$d$ valence electrons. This causes RIXS to couple to multi-magnon excitations, as was first pointed out in Ref.~\onlinecite{Brink05b}.  The simplest microscopic mechanism for this coupling is obtained within the strong-coupling Hubbard model, in which the doubly occupied and empty virtual states shift in energy in presence of the core-hole~\cite{Brink05b,Nagao07}. Adding the amplitudes for the two possible processes shown in Fig.~\ref{fig:alteredJ}, lead to an exchange integral in presence of a core-hole on site $i$ of
\begin{equation}
    J^c_{ij} = \frac{2t^2_{ij}}{U+U_c} + \frac{2t^2_{ij}}{U-U_c} = J_{ij} \left( 1 + \eta \right) \label{eq:Jc}
\end{equation}
where $j$ is a site neighboring to $i$ and $\eta = U^2_c/(U^2-U^2_c)$. This enables us to write down the generic Hamiltonian for the intermediate states~\cite{Brink05b}:
\begin{equation}
    H_{\text{int}} = H_0 + \eta \sum_{i,j} s^{\phantom{\dag}}_i s^{\dag}_i J_{ij} {\bf S}_i \! \cdot \! {\bf S}_j\label{eq:Hint}
\end{equation}
where $s_i$ creates a core-hole and $s^{\dag}_i$ annihilates one at site $i$. In the Hubbard framework one could identify the $U$ with the Coulomb energy associated with two holes in a $3d$-orbital $U_d = 8.8$ eV, which together with $U_c = 7.0$ eV~\cite{Barriquand94,Okada06} leads to $\eta=1.7$; from $U/U_c =2/3$, as suggested in Ref.~\onlinecite{Tsutsui99}, one finds $\eta=-0.8$.

\begin{figure}
\begin{center}
\includegraphics[width=\columnwidth]{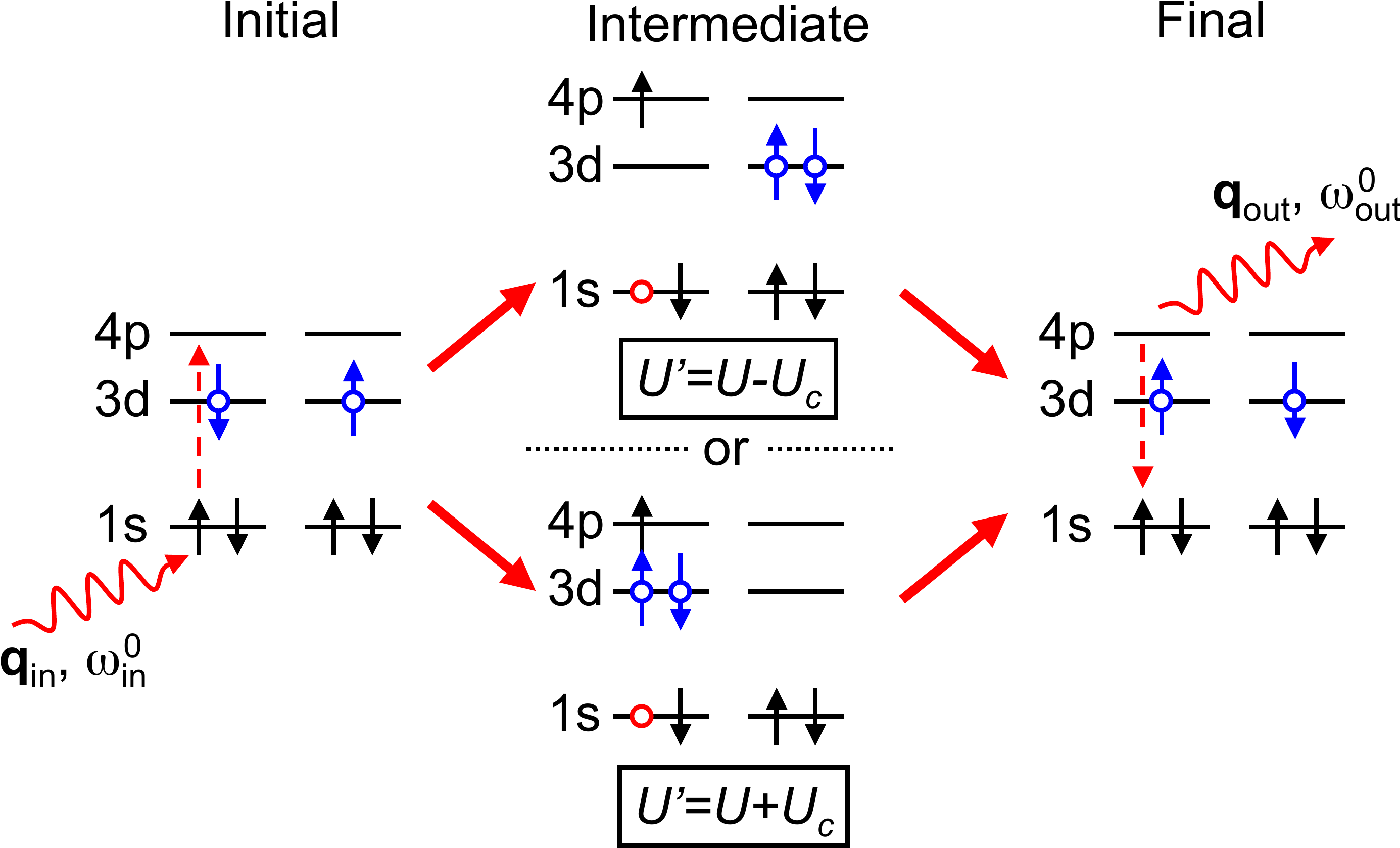}
\end{center}
\caption{In RIXS, a photon of momentum ${\bf q}_{\text{in}}$ and energy tuned to the $K$-edge of a transition metal ion $(\omega^0_{\text{in}} = \omega_{\text{res}})$ creates a core-hole at a certain site. The superexchange interaction between this site and a neighboring other site is modified because the energy of the virtual intermediate states is changed. The same-site Coulomb repulsion $U$ is lowered by $U_c$ if the core-hole site contains no holes and is raised by $U_c$ if there are two holes present. Summing the amplitudes for both processes, we obtain the modified superexchange interaction, see Eq.~(\ref{eq:Jc}).\label{fig:alteredJ}}
\end{figure}

The situation in the cuprates, however, is more complex and one needs to go beyond the single band Hubbard model to obtain a value of $\eta$ from microscopic considerations. We will do so by considering a three-band model in the strong coupling limit. However, it should be emphasized that for the end result --the computed RIXS spectrum in the UCL approach-- $\eta$ just determines the overall scale of the inelastic scattering intensity. As we will show higher order corrections in the UCL approach are determined by the value of $\eta$, because $\eta J/\Gamma$ appears as a small parameter in this expansion. As for the cuprates $J/\Gamma \approx 1/5$ such corrections are small for the relevant possible values of $\eta$.

In the three-band Hubbard model that includes also the oxygen states, two important kinds of intermediate states appear: the poorly- and well-screened ones. Because the Coulomb interaction of the core-hole with the valence electrons is large ($U_c=7.0$ eV, compared to a charge transfer energy $\Delta = 3.0$ eV\cite{Okada06}), a copper hole can transfer to a neighboring oxygen to form a well-screened intermediate state. The low-energy sector now also encompasses an oxygen hole, equally distributed over the ligands. We will show that, starting from a three band Hubbard model, Eq.~(\ref{eq:Hint}) gives a proper description of both the well- and poorly-screened intermediate states, with $\eta$ now a function of the parameters of the three band model. Before presenting these results we remark that scattering processes that scatter a well-screened state into a poorly-screened state or vice versa yield a large energy loss $\omega$. These are not important at low $\omega$, where one will only observe scattering in the magnetic channel, not the charge one. 

\begin{figure}[!htp]
	\begin{center}
	\includegraphics[width=\columnwidth]{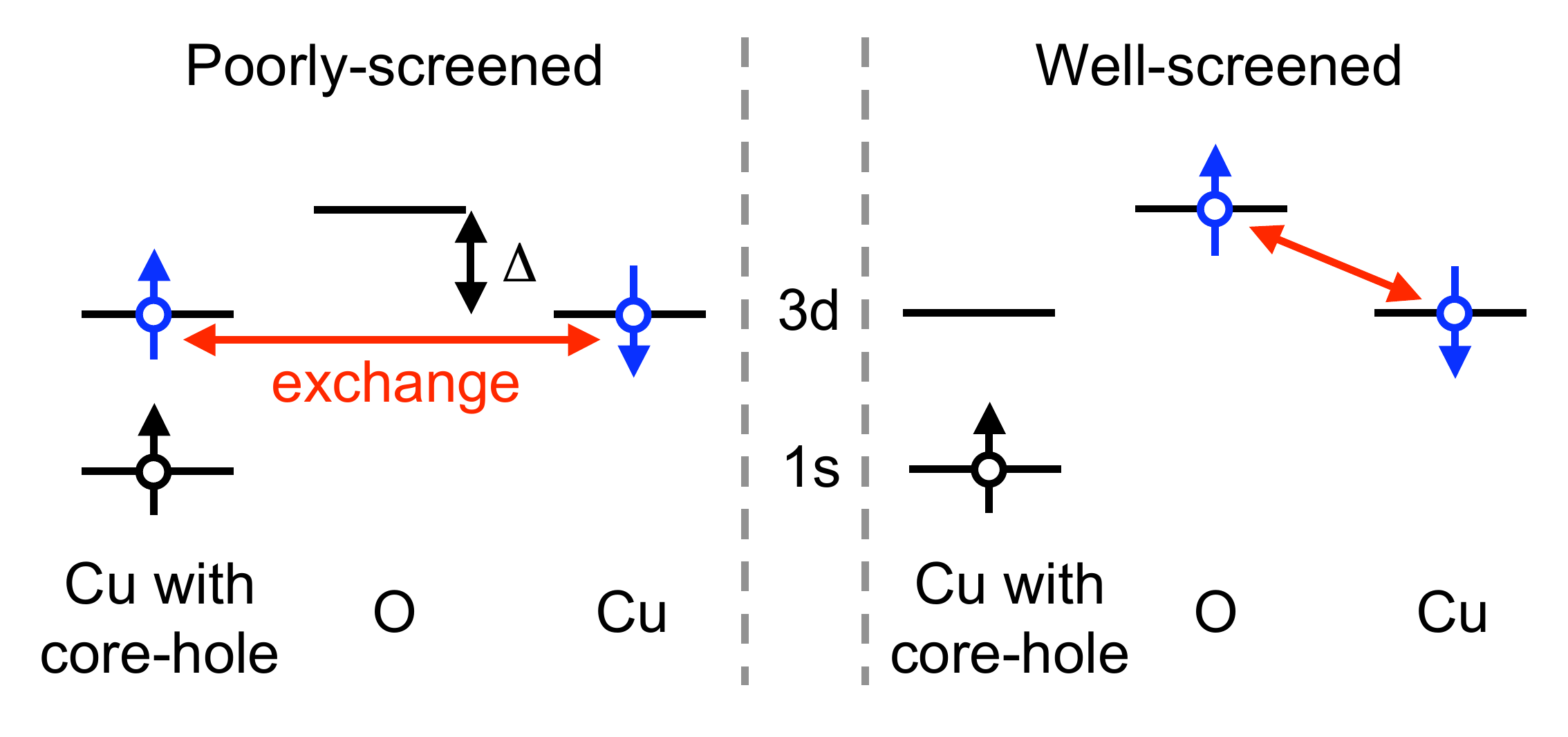}
	\end{center}
	\caption{Modification of the superexchange interaction in the well- and poorly-screened intermediate states. In the poorly-screened state, the core-hole potential $U_c$ modifies the superexchange. For the well-screened state however, the copper 3$d$ hole on the core-hole site is transfered to a neighboring oxygen, and superexchange is only of order $\mathcal{O}(t_{pd}^2)$, independent of $U_c$.\label{fig:threeband}}
\end{figure}

The magnetic scattering processes for the poorly-screened state are very similar to the single band picture: all copper ions have one hole and all oxygen ions are filled-shell. The superexchange processes are shown in Fig.~(\ref{fig:threeband}). We consider the Anderson and Geertsma contributions to the superexchange~\cite{Barriquand94} and find
\begin{align}
	\eta_{ps} &= \frac{U_d \Delta^2 (U_p+2\Delta)}{2(2U_d+2\Delta+U_p)}\left( \frac{1}{(U_d-U_c)(U_c-\Delta)^2} \right. \nonumber \\
	&\left. +\frac{1}{(U_d+U_c)\Delta^2}+\frac{\left[1/\Delta  + 1/(\Delta-U_c) \right]^2}{2\Delta-U_c+U_p} \right)-1, 
\end{align}
which results in $\eta=-0.3$ using the parameters $U_d = 8.8$ eV, $U_p = 6.0$ eV, $t_{pd} = 1.3$ eV, $\Delta = 3.0$ eV, and $U_c = 7.0$ eV\cite{Barriquand94,Okada06}, where $t_{pd}$ is the copper-oxygen hoping integral and $U_p$ the on-site Coulomb repulsion of two oxygen holes. 

The well-screened intermediate states have a similarly modified superexchange interaction, as shown in Fig.~\ref{fig:threeband}. Because of the large core-hole Coulomb interaction an electron from the neighboring oxygen atoms moves in to screen it, or, equivalently, the copper hole is transferred to the in-plane oxygen ions. Transfer out of the plane is not considered since the Cu 3$d_{x^2-y^2}$ hole only couples to the in-plane oxygens. Because the Cu hole is transfered in the direction of one of its neighboring Cu ions, the contribution to the superexchange interaction for the well-screened state is of second order in $t_{pd}$, instead of fourth order between two Cu sites (see Fig.~\ref{fig:threeband}). The rotational invariance around the core-hole site of the transfered hole ensures that the intermediate state Hamiltonian of the form Eq.~(\ref{eq:Hint}) gives the correct scattering amplitude. To lowest order in $t_{pd}$ we hence find
\begin{equation}
	\eta_{ws} = \frac{U_d(U_d+U_p)\Delta^2(U_p+2\Delta)}{2 (U_d-\Delta) t_{pd}^2 (2U_d + U_p + 2\Delta)(U_p+\Delta)} - 4, 
\end{equation}
which results in $\eta=-1.3$ --again restricting ourselves to superexchange of the Anderson and Geertsma type. We see that to lowest order, the core-hole potential $U_c$ does not appear in the well-screened intermediate state. From these microscopic considerations we conclude that the intermediate state Hamiltonian Eq.~(\ref{eq:Hint}) is the correct one and higher order corrections to it are small because for the cuprates $\eta$ is a number of order unity.

In a previous theoretical treatment we have shown in detail how to derive the cross section for RIXS-processes with a local core-hole using the UCL expansion~\cite{Ament07}. For an incoming/outgoing photon with momentum ${\bf q}_{\text{in}}/{\bf
q}_{\text{out}}$ and energy $\omega^0_{\text{in}}/\omega^0_{\text{out}}$, we obtained the
cross section through the Kramers-Heisenberg
relation~\cite{Kramers25,Platzman69,Klein83,Blume85} as a function
of energy loss $\omega = \omega^0_{\text{in}} -
\omega^0_{\text{out}}$ and momentum transfer ${\bf q} = {\bf
q}_{\text{out}} - {\bf q}_{\text{in}}$:
\begin{align}
    \left. \frac{d^2\sigma}{d\Omega d\omega} \right|_{\text{res}} &\propto \left< \sum_f \left| A_{fi} \right|^2 \delta (\omega - \omega_{fi}) \right>_T\!\!, \;\;\;\text{with} \label{eq:kramersheisenberg} \\
    A_{fi} &= \omega_{\text{res}} \sum_n \frac{\bok{f}{\hat{D}}{n}\bok{n}{\hat{D}}{i}}{\omega_{\text{in}}-E_n - i \Gamma}.
\end{align}
The initial state $\ket{i}$ with energy $E_i$ (which is used as reference energy: $E_i=0$) is photo-excited to an intermediate state which is described by the dipole operator $\hat{D}$. The system can evolve through the intermediate states $\ket{n}$ with energy $E_n$ (measured with respect to the resonance energy $\omega_{\text{res}}$) and, after the decay of the core-hole, end up in a final state $\ket{f}$ with energy $E_f$. Because the life time of the core-hole is ultrashort, we introduce an energy broadening $\Gamma$ for the intermediate state. The detuning of the incoming photon energy from the $K$-edge is given by $\omega_{\text{in}} = \omega^0_{\text{in}} - \omega_{\text{res}}$. Finally, the delta function in Eq.~(\ref{eq:kramersheisenberg}) imposes energy conservation: the energy gain of the system $\omega_{fi} = E_f - E_i$ must be equal to the energy loss of the photon $\omega = \omega^0_{\text{in}} - \omega^0_{\text{out}}$. If $\Gamma > E_n$ we can expand the amplitude $A_{fi}$ in a powerseries. We assume that the energy of the incoming photon is tuned to the resonance ($\omega_{\text{in}} = 0$):
\begin{equation}
    A_{fi} = \frac{\omega_{\text{res}}}{- i \Gamma} \sum^{\infty}_{l=1} \frac{1}{(- i \Gamma )^l} \bra{f} \hat{D} (H_{\text{int}})^l \hat{D} \ket{i}.
\end{equation}
Note that we left out the $l=0$ term because it only contributes to elastic scattering. The leading order non-vanishing term in the sum is $l=1$, since the core-hole broadening is quite large compared to $J$. At the copper $K$-edge is $2\Gamma \approx 1.5$ eV according to Refs.~\onlinecite{Krause79,Hamalainen89}, and $2\Gamma \approx 3$ eV for the closely related ions Mn and Ge according to Refs.~\onlinecite{Shen06,Elfimov02}, which in either case is large compared to $J$.  As in the three-band model  $\eta = -1.3/-0.3$ eV for the well-/poorly-screened intermediate state, the largest value we find is $\eta J/\Gamma \approx -0.22$. Note that the UCL expansion therefore converges very well --even faster for the poorly-screened state than for the well-screened state (where $|\eta|$ is larger). It is possible to directly include a number of terms with $l \ge 2$ in the cross section by using the expansion
\begin{equation}
    \sum_{l=1}^{\infty} \frac{(H_{\text{int}})^l}{\Gamma^l} \approx \sum_{l=1}^{\infty} \left( \frac{H^l_0}{\Gamma^l} + \frac{H^{l-1}_0 H'}{\Gamma^l} \right) + \mathcal{O} \bigl( (\eta J/\Gamma)^2 \bigr) \label{eq:Hintapprox}
\end{equation}
with $H' = \eta \sum_{i,j} s^{\phantom{\dag}}_i s^{\dag}_i J_{ij} {\bf S}_i \! \cdot \! {\bf S}_j$. Since $[H_0,\hat{D}] = 0$ and $H_0 \ket{i} = 0$, all terms with $H_0$ on the right can be safely neglected. Using Eq.~(\ref{eq:Hintapprox}), $A_{fi}$ simplifies to
\begin{equation}
    A_{fi} = \frac{\omega_{\text{res}}}{i \Gamma} \frac{\eta}{i \Gamma + \omega} \bra{f} \hat{O}_{\bf q} \ket{i}\label{eq:Afi}
\end{equation}
with the scattering operator
\begin{equation}
    \hat{O}_{\bf q} = \sum_{i,j} e^{i {\bf q} \cdot {\bf R}_i} J_{ij} {\bf S}_i \!\cdot\! {\bf S}_j.\label{eq:Orealspace}
\end{equation}
From this equation we can deduce two important features. Firstly, indirect RIXS probes a momentum dependent four-spin correlation function\cite{Brink05b}. Secondly, $\hat{O}_{\bf q}$ commutes with the $z$-component of total spin $S_z$, so the allowed scattering processes should leave $S_z$ unchanged. Only an even number of magnons can be created or annihilated.

To bosonize Eq.~(\ref{eq:Orealspace}), we split $\hat{O}_{\bf q}$ in four parts:
\begin{equation}
    \hat{O}_{\bf q} = \sum_{i,j\in A} \dots + \sum_{i,j\in B} \dots + \sum_{i\in A,\; j\in B} \dots + \sum_{i\in B,\; j\in A} \dots
\end{equation}
Next, we rewrite this expression using LSWT as introduced in section \ref{sec:general}. Fourier transforming the result gives
\begin{widetext}
    \begin{align}
        \hat{O}_{\bf q} = \text{const} + S \sum_{\bf k} &\left[ \left( J^{AA}_{{\bf k} + {\bf q}/2} + J^{AA}_{{\bf k} - {\bf q}/2} - J^{AA}_{\bf 0} - J^{AA}_{\bf q} + J^{AB}_{\bf 0} + J^{AB}_{\bf q} \right) \left( a^{\dag}_{{\bf k} - {\bf q}/2} a^{\phantom{\dag}}_{{\bf k} + {\bf q}/2} + b^{\dag}_{{\bf k} - {\bf q}/2} b^{\phantom{\dag}}_{{\bf k} + {\bf q}/2} \right) + \right. \nonumber \\
        &\;\;\left. \left( J^{AB}_{{\bf k} + {\bf q}/2} + J^{AB}_{{\bf k} - {\bf q}/2} \right) \left( a^{\phantom{\dag}}_{{\bf k} + {\bf q}/2} b^{\phantom{\dag}}_{-{\bf k} + {\bf q}/2} + a^{\dag}_{{\bf k} - {\bf q}/2} b^{\dag}_{-{\bf k} - {\bf q}/2} \right) \right]
    \end{align}
\end{widetext}
and we can write $\hat{O}_{\bf q}$ in terms of the magnon operators using the inverses of Eqs.~(\ref{eq:alpha}) and (\ref{eq:beta}). This leads to
\begin{equation}
    \hat{O}_{\bf q} = \hat{O}^{(1)}_{\bf q} + \hat{O}^{(2)}_{\bf q}\label{eq:O}
\end{equation}
where $\hat{O}^{(1,2)}_{\bf q}$ is a lengthy expression that contains the one/two-magnon scattering part. The next section deals with the two-magnon part $\hat{O}^{(2)}_{\bf q}$ where two magnons are created or annihilated. The one-magnon part $\hat{O}^{(1)}_{\bf q}$ (where the change in the number of magnons is zero) is treated in section \ref{sec:T>0}.
\end{section}

\begin{section}{Two-Magnon Scattering at $T=0$ K\label{sec:T=0}}
At $T=0$ K, the system is in its groundstate, where no magnons are present: $\ket{i} = \ket{0}$. Adding conservation of $S_z$, the only allowed scattering processes are the ones in which two magnons are created, so we consider the two-magnon part of the scattering operator of Eq.~(\ref{eq:O}) with $S=1/2$:
\begin{widetext}
\begin{align}
    \hat{O}^{(2)}_{\bf q} = \sum_{{\bf k} \in MBZ} &\biggl[ -\left( J^{AA}_{{\bf k}+{\bf q}/2} + J^{AA}_{{\bf k}-{\bf q}/2} - J^{AA}_{\bf 0} - J^{AA}_{\bf q} + J^{AB}_{\bf 0}+J^{AB}_{\bf q} \right) \left(u_{{\bf k}+{\bf q}/2} v_{{\bf k}-{\bf q}/2} + u_{{\bf k}-{\bf q}/2} v_{{\bf k}+{\bf q}/2} \right) + \biggr.\nonumber \\
    &\biggl. \left( J^{AB}_{{\bf k}+{\bf q}/2} + J^{AB}_{{\bf k}-{\bf q}/2} \right) \left( u_{{\bf k}+{\bf q}/2} u_{{\bf k}-{\bf q}/2} + v_{{\bf k}+{\bf q}/2} v_{{\bf k}-{\bf q}/2} \right) \biggr] \left( \alpha_{{\bf k}+{\bf q}/2} \beta_{-{\bf k}+{\bf q}/2} + \alpha^{\dag}_{{\bf k}-{\bf q}/2} \beta^{\dag}_{-{\bf k}-{\bf q}/2} \right)
\end{align}
\end{widetext}
The two-magnon spectrum is shown in Fig.~\ref{fig:rixsdos}(a). Several remarkable features can be seen.

\begin{figure}
\begin{center}
\begin{tabular}{c}
\begin{minipage}{\columnwidth}
\hspace{-1cm}\includegraphics*[width=\columnwidth]{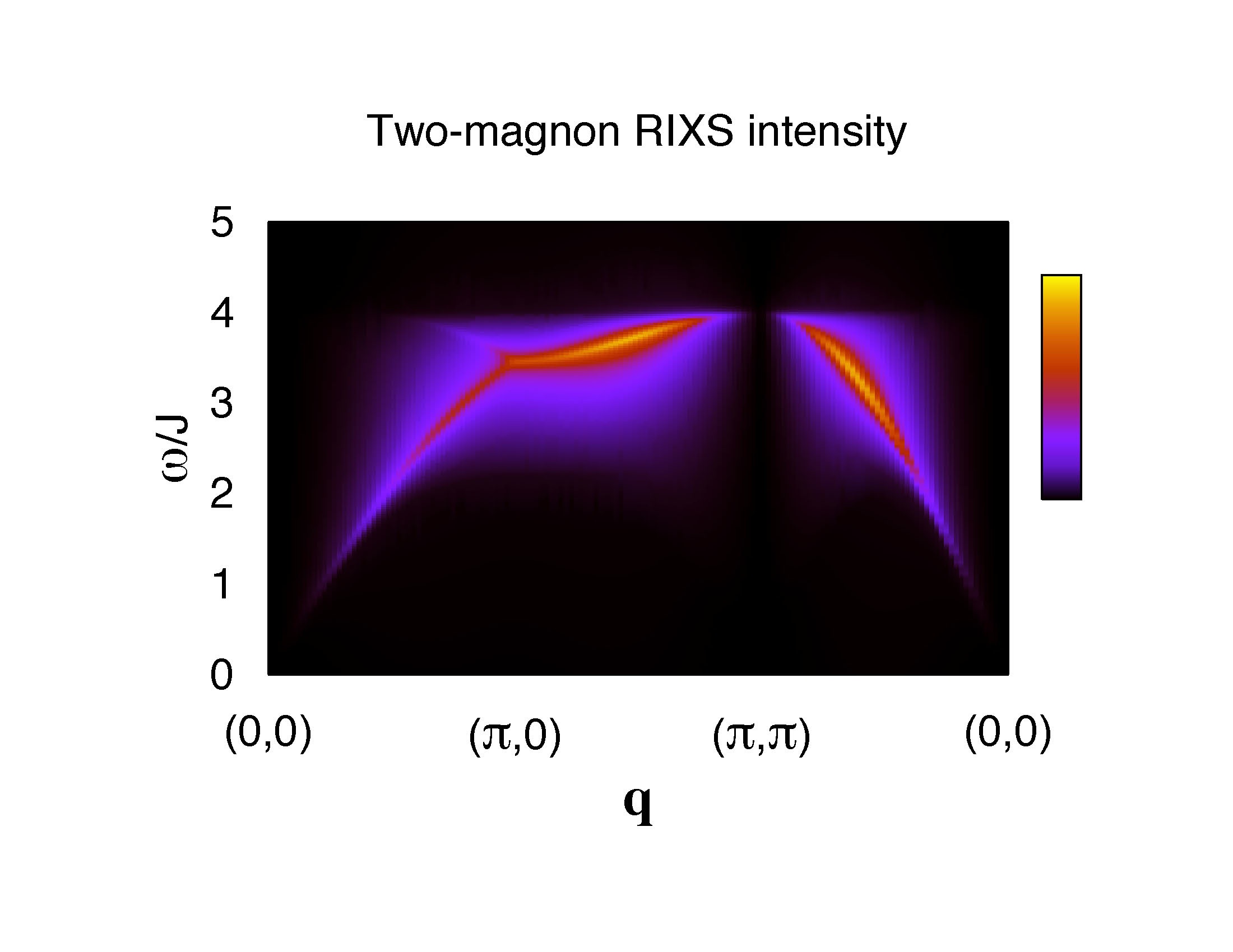}\vspace{-1cm}
\begin{center}\hspace{-0.5cm}(a)\end{center}
\end{minipage}
\\
\begin{minipage}{\columnwidth}
\hspace{-1cm}\includegraphics[width=\columnwidth]{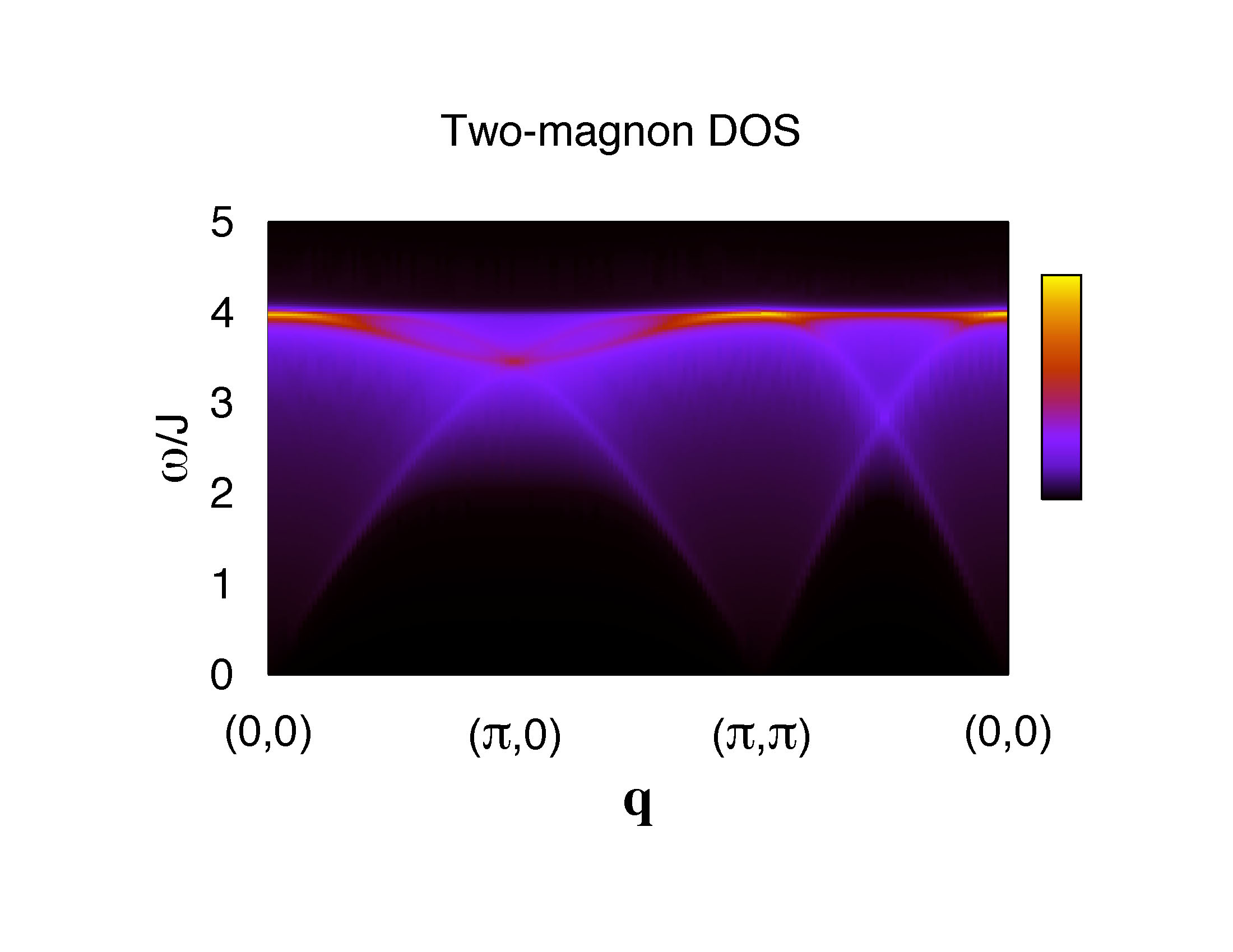}\vspace{-1cm}
\begin{center}\hspace{-0.5cm}(b)\end{center}
\end{minipage}
\\
\begin{minipage}{\columnwidth}
\hspace{0.3cm}\includegraphics[angle=270,width=0.5\columnwidth]{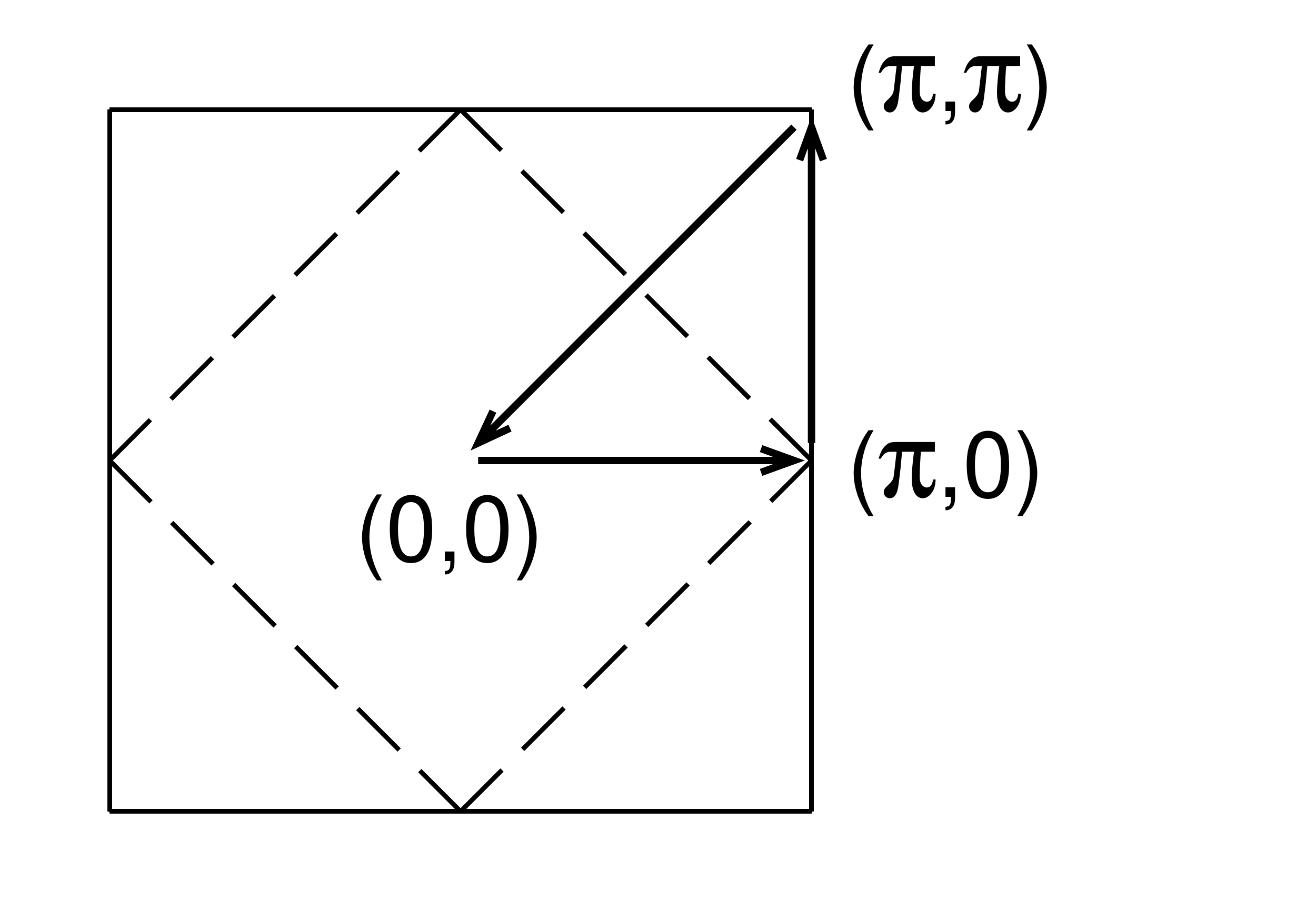}
\begin{center}\hspace{-0.5cm}(c)\end{center}\vspace{-.7cm}
\end{minipage}
\end{tabular}
\end{center}
\caption{RIXS spectrum (a) and two-magnon DOS (b) for a nearest neighbor Heisenberg antiferromagnet with exchange interaction $J$ as a function of transferred momentum {\bf q} for a cut through the  Brillouin zone (c). The dashed line indicates the magnetic BZ boundary.\label{fig:rixsdos}}
\end{figure}

\begin{figure}
\begin{center}
\begin{tabular}{c}
\begin{minipage}{\columnwidth}
\hspace{-.6cm}\includegraphics[angle=270,width=0.75\columnwidth]{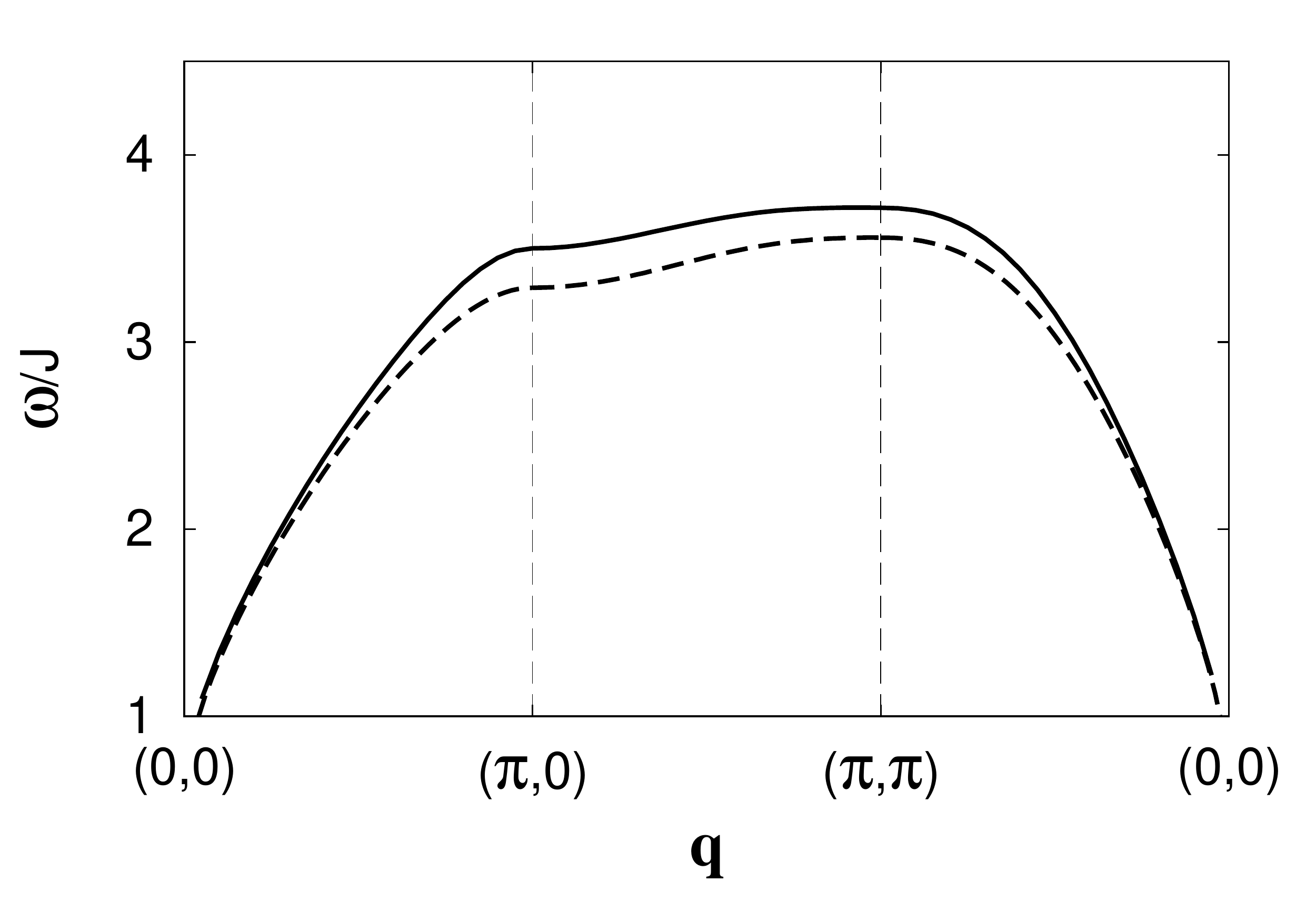}
\begin{center}(a)\end{center}
\end{minipage}
\\
\begin{minipage}{\columnwidth}
\includegraphics[angle=270,width=0.7\columnwidth]{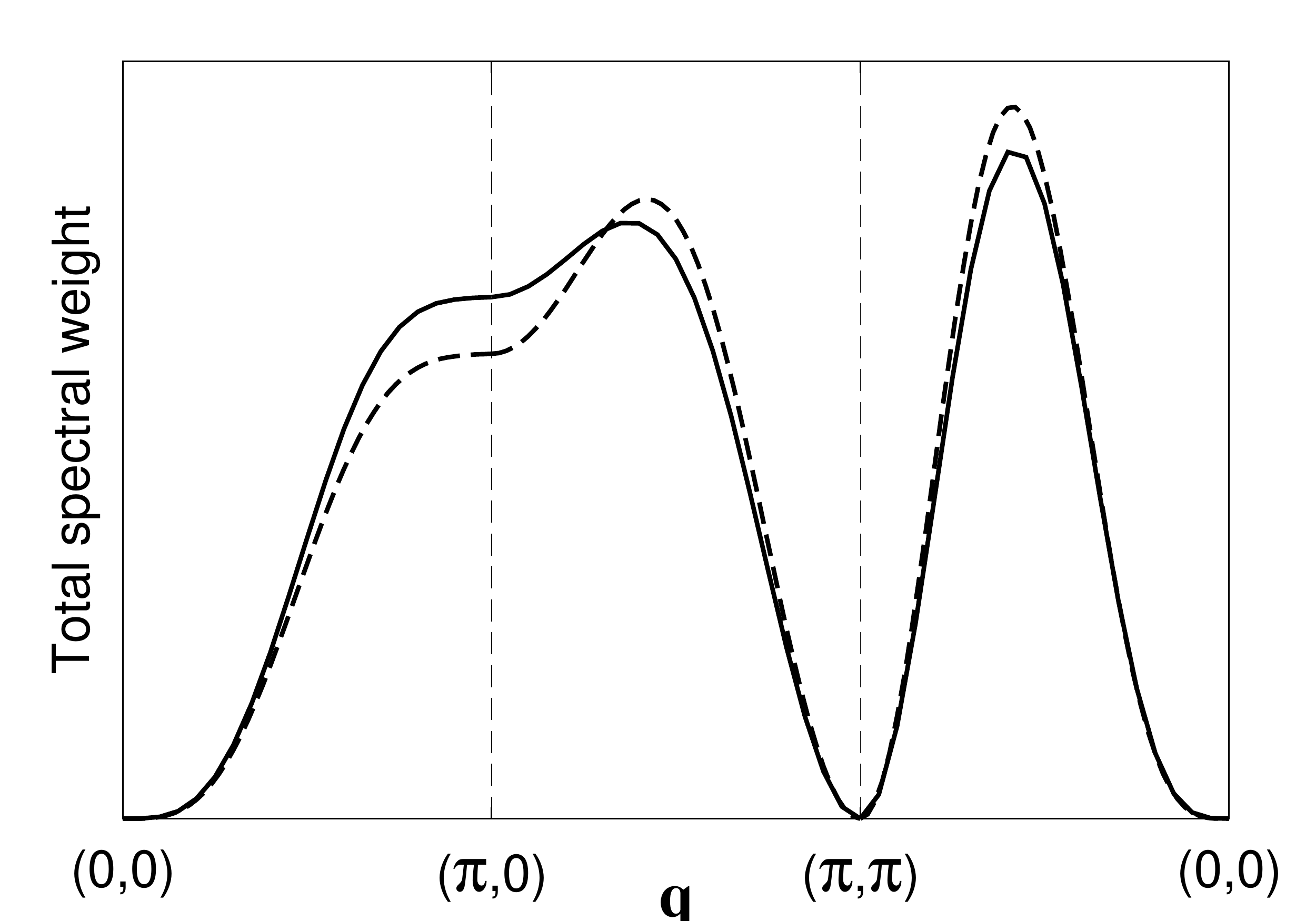}
\begin{center}(b)\end{center}
\caption{First moment (a) and total spectral weight (b) of the RIXS spectrum. The solid lines are
obtained by using interaction strengths determined from neutron
data (next neighbor coupling $J=146.3$ meV, second and third
neighbor couplings $J'=J''=2 $ meV and ring exchange $J_c = 61$
meV).~\cite{Coldea01} The dashed lines have only nearest neighbor
interaction.\label{fig:momentweight}}
\end{minipage}
\end{tabular}
\end{center}
\end{figure}

First of all the spectral weight vanishes at ${\bf q} = (0,0)$ and ${\bf q} = (\pi,\pi)$, as can be seen in Fig.~\ref{fig:momentweight}(b). This is in agreement with experimental observations~\cite{Hill_tbp}. The vanishing of the RIXS intensity at ${\bf q} = {\bf 0}$ is obvious: from Eq.~(\ref{eq:Orealspace}) we see that at ${\bf q} = {\bf 0}$, $\hat{O}_{\bf q}$ reduces to $2H_0$ (the factor of $2$ arises from the fact that the sum in Eq.~(\ref{eq:Orealspace}) is over all $i$ and $j$). At zero temperature, $\ket{i} = \ket{0}$ and consequently $H_0 \ket{i} = 0$ --the RIXS intensity vanishes. At nonzero temperatures, $H_0 \ket{i} = E_i \ket{i}$ and according to Eq.~(\ref{eq:Afi}) only elastic scattering occurs. It is easy to show that at ${\bf q} =(\pi,\pi)$ the RIXS intensity always vanishes, regardless of the temperature or the form of $J_{ij}$ (as long as there is antiferromagnetic order). This holds because ${\bf q} = (\pi,\pi)$ is a reciprocal magnetic lattice vector: $e^{i {\bf q}\cdot {\bf R}_i} = 1$ if ${\bf R}_i$ is in sublattice $A$ and $e^{i {\bf q}\cdot {\bf R}_i} = -1$ if ${\bf R}_i$ is in sublattice $B$ (assuming that at ${\bf R}_i = (0,0)$ we are in sublattice $A$).
We obtain
\begin{equation}
    \hat{O}_{{\bf q}=(\pi,\pi)} = \sum_{i \in A,j} J_{ij} {\bf S}_i \! \cdot \! {\bf S}_j - \sum_{i \in B,j} J_{ij} {\bf S}_i \! \cdot \! {\bf S}_j.
\end{equation}
Adding all terms where $j \in B$ in the first term and $j \in A$ in the latter, we get zero. What remains is
\begin{equation}
    \hat{O}_{{\bf q}=(\pi,\pi)} = \sum_{i,j \in A} J_{ij} {\bf S}_i \! \cdot \! {\bf S}_j - \sum_{i,j \in B} J_{ij} {\bf S}_i \! \cdot \! {\bf S}_j.
\end{equation}
These terms cancel when applied to an initial state which is symmetric under the interchange of the sublattices.

The other remarkable feature of the magnetic RIXS spectrum is its strong dispersion. This is apparent from Fig.~\ref{fig:rixsdos}(a) and \ref{fig:momentweight}(a), showing the first moment (average peak position) of the spectrum. The calculations for the nearest neighbor Heisenberg antiferromagnet (see the dashed line in Fig.~\ref{fig:momentweight}(a)) show that the magnetic scattering disperses from about $\omega \approx 0$ around $(0,0)$ to $\omega \approx 4
J$ at $(\pi,0)$ and $(\pi/2,\pi/2)$. Longer range couplings tend to reduce (increase) the first moment of the RIXS spectrum if they weaken (reinforce) the antiferromagnetic order (see the solid line in Fig.~\ref{fig:momentweight}(a)). The observed dispersion in Fig.~\ref{fig:rixsdos}(a) has a two-fold origin. It is in part due to the {\bf q}-dependence of the
two-magnon density of states (DOS), combined with the scattering matrix elements that tend to pronounce the low energy tails of the two-magnon DOS. In Fig.~\ref{fig:rixsdos}(b), it looks as if the two-magnon DOS has two branches. The most energetic one around ${\bf q} = {\bf 0}$ is strongly suppressed by the matrix elements throughout the Brillouin zone (BZ).

The consistency at ${\bf q} = (0,0)$ and ${\bf q} =(\pi,\pi)$ of the theoretical results and experimental data was already noticed, but at other wave-vectors the agreement stands out even more. The data on La$_2$CuO$_4$ for ${\bf q} = (\pi,0)$ shows a peak at around $500$ meV, precisely where we find it on the basis of a nearest neighbor Heisenberg model with $J=146$ meV -- a value found by the analysis of neutron scattering data~\cite{Coldea01}. Similar agreement is found at ${\bf q} = (0.6\pi,0)$ and ${\bf q} = (0.6\pi,0.6\pi)$.\cite{Hill_tbp} Even better agreement is found when we take into account the second and third nearest neighbors and ring exchange according to the neutron data. The ring exchange interaction, which we treat on a mean field level, simply renormalizes first- and second-nearest neighbors exchange~\cite{Coldea01}.

In Fig.~\ref{fig:compare}, we compare the results for the two-magnon scattering intensity with experimental data,\cite{Hill_tbp} using the interaction strengths determined from neutron data~\cite{Coldea01}, for three values of
${\bf q}$ in the BZ. Note that we use the wave-vector independent renormalization factor $Z_c$ here, that takes into account some of the magnon-magnon interactions.~\cite{Oguchi60} This simply changes the energy scale by a factor $Z_c \approx 1.18$ but does not affect the intensity of the spectrum. Each panel shows the theoretical prediction (dashed line), the theory convoluted with the current instrumental resolution (solid line), and the experimental data. The only free parameter in the theoretical spectra is the over-all scale of the scattering intensity. We find it to vary by a factor of $2.5$ comparing different ${\bf q}$'s, which is within the error bars of the experiment~\cite{Hill_pc}.

Many qualitative features such as the occurrence of intense peaks
at the magnetic BZ boundary and the large dispersion
characterizing the total spectrum are in accordance with our
earlier results\cite{Brink05b} and the results of Nagao and
Igarashi~\cite{Nagao07}. The spectra of Ref. \onlinecite{Nagao07},
taking two-magnon interactions partially into account, show slight
quantitative differences with respect to our results: the RIXS
peaks soften and broaden somewhat as a consequence of the
magnon-magnon interaction, particularly for the ($\pi$,0) point.
The range of the dispersion in the spectrum is therefore smaller
(the mean $\omega/J$ varies between 1 and 3 instead of 1 and 4).

\begin{figure}[tp]
\begin{center}
\begin{tabular}{cc}
\begin{minipage}{0.5\columnwidth}
\hspace{-1cm}
\includegraphics*[angle=270,width=1.1\columnwidth]{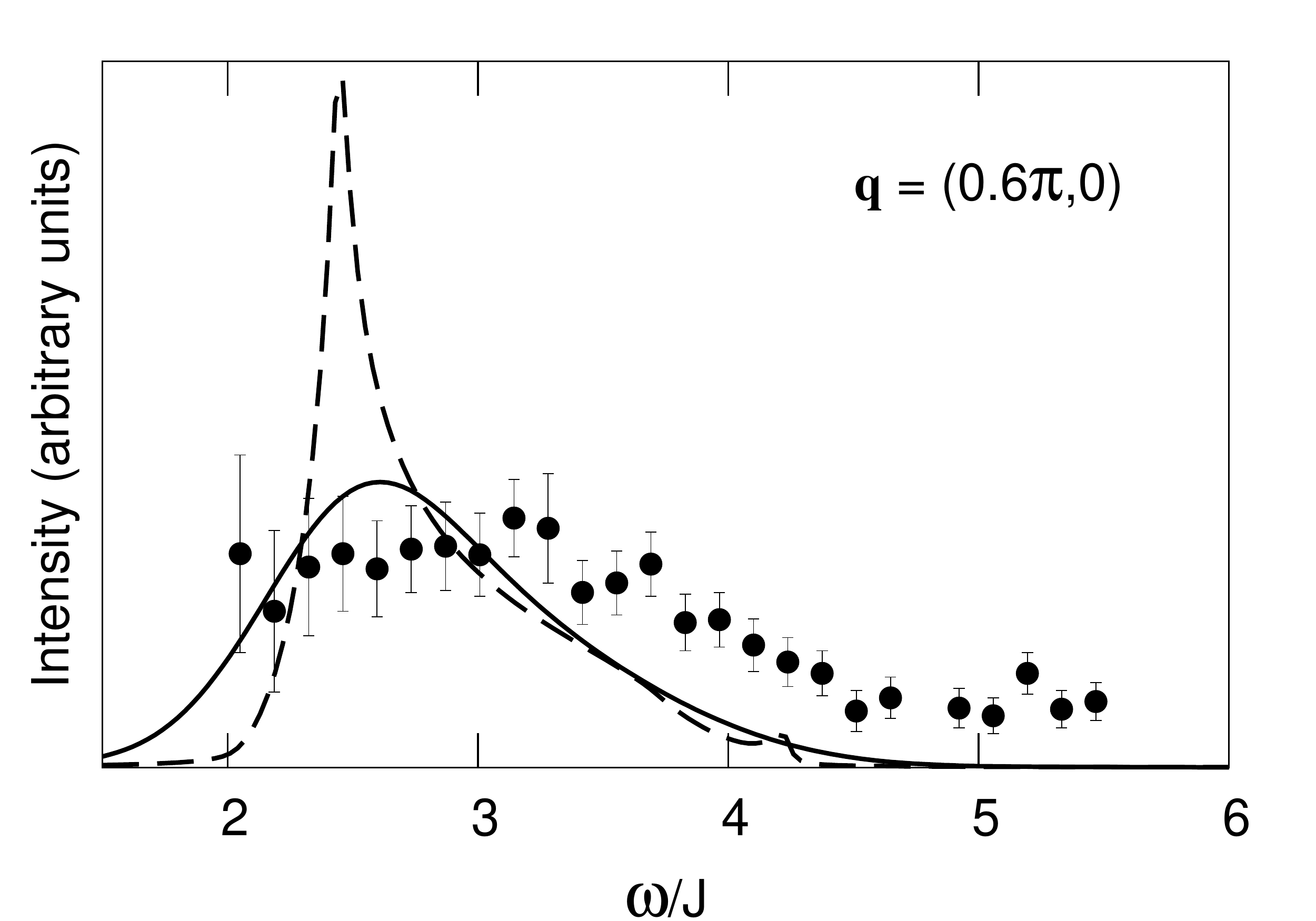}
\begin{center}(a)\end{center}
\end{minipage}
&
\begin{minipage}{0.5\columnwidth}
\hspace{-1cm}
\includegraphics*[angle=270,width=1.1\columnwidth]{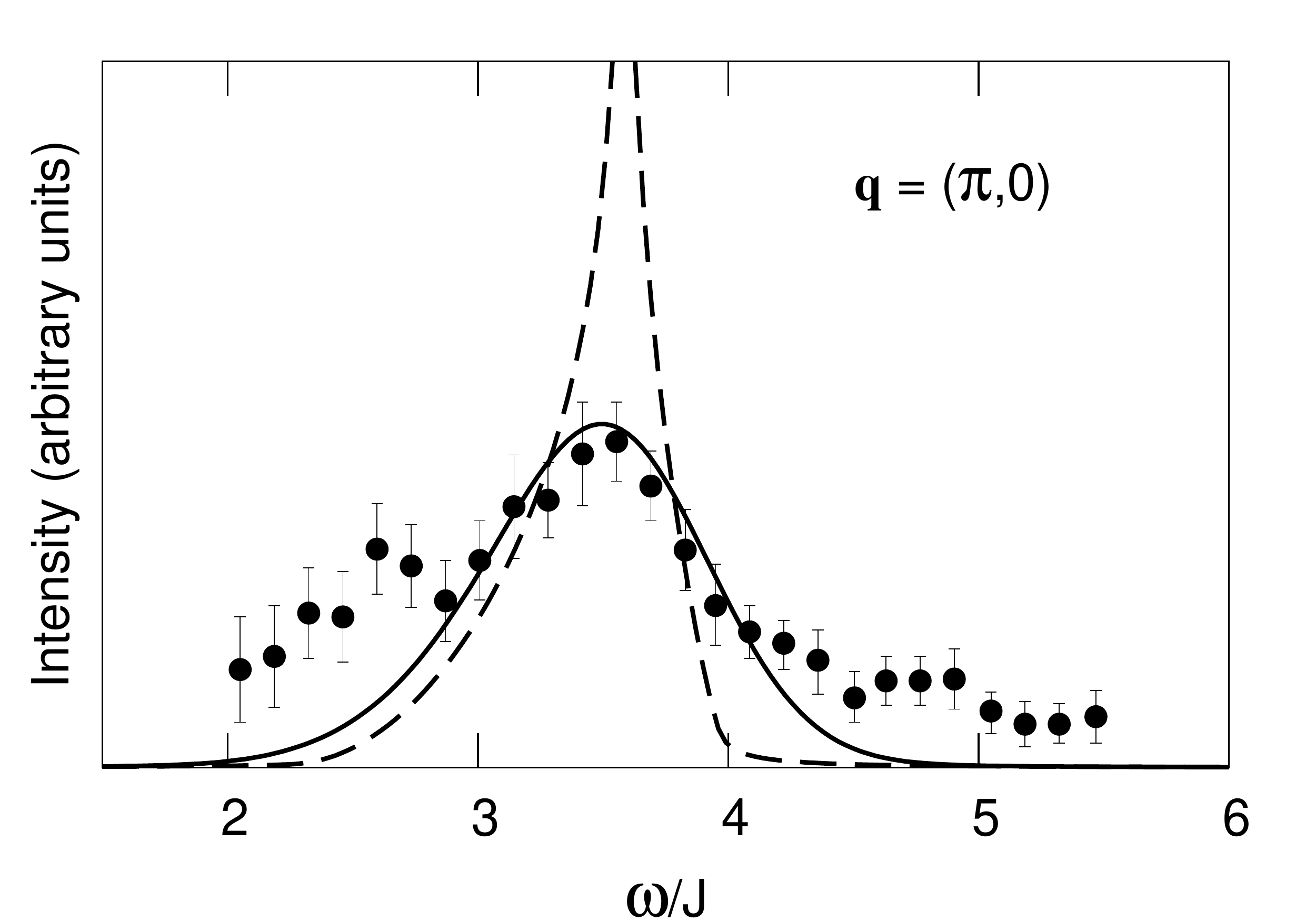}
\begin{center}(b)\end{center}
\end{minipage}
\\
\begin{minipage}{0.5\columnwidth}
\hspace{-1cm}
\includegraphics*[angle=270,width=1.1\columnwidth]{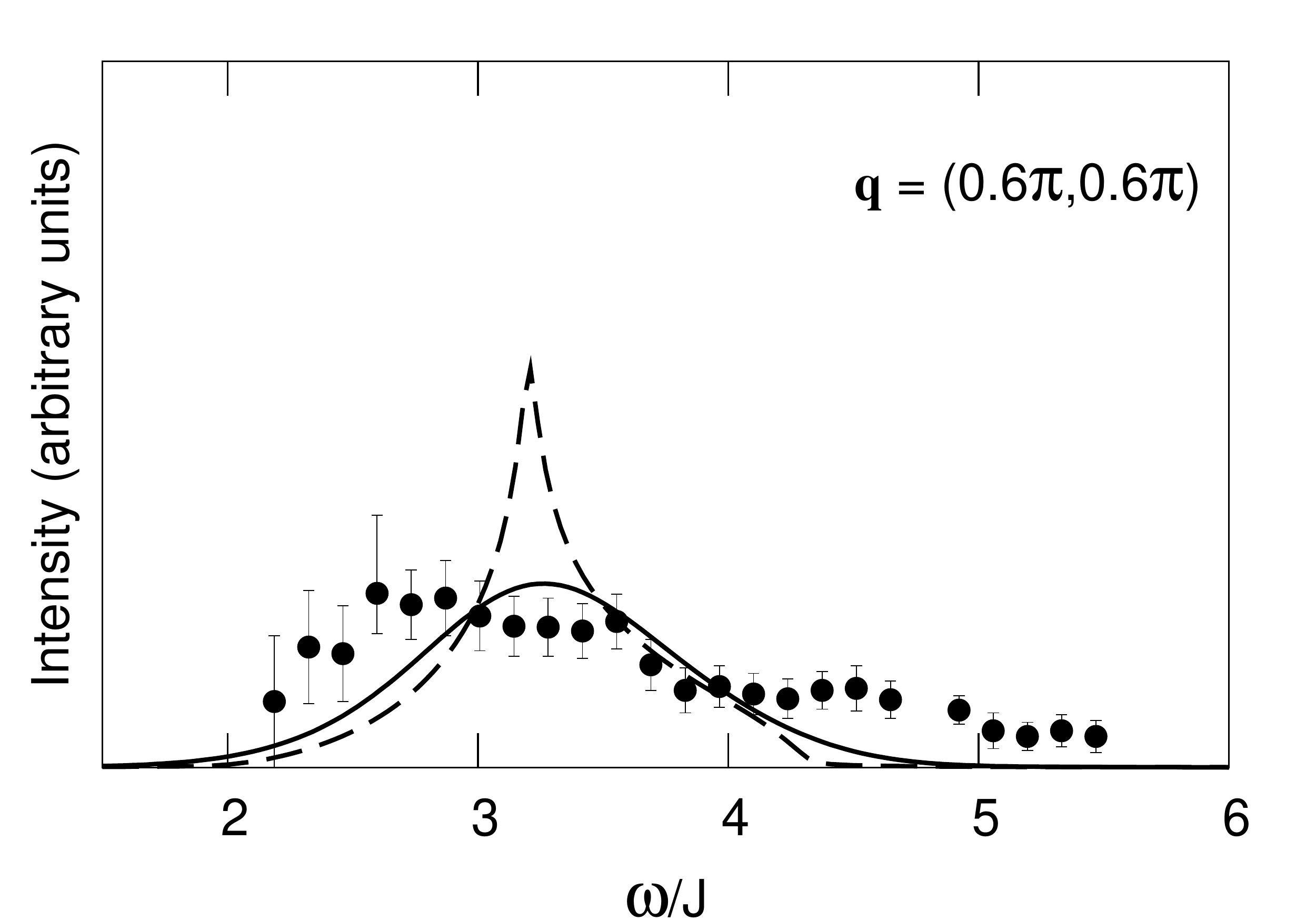}
\begin{center}(c)\end{center}
\vspace{-.8cm}
\end{minipage}
&
\end{tabular}
\end{center}
\caption{RIXS intensity for various points in the BZ. Each figure
contains the bare theoretical data (dashed line), the convolution
with experimental resolution (solid line), and the experimental
data from Ref.~\onlinecite{Hill_tbp}. For these figures, we used
$J=146.3$ meV, second and third neighbor couplings $J'=J''=2 $ meV
and ring exchange $J_c = 61$ meV. The latter contribution is
evaluated theoretically using a mean field approximation. These
values were found in neutron scattering
experiments.\cite{Coldea01} These experiments were analyzed using
the wave-vector independent renormalization factor $Z_c = 1.18$,
which is also used to generate the theoretical curves. The
theoretical intensity is scaled independently in each figure to
match the experiment. The overall scale factors differ at most by a factor $2.5$, which is comparable to experimental
uncertainty in absolute
intensities.\cite{Hill_pc}\label{fig:compare}}
\end{figure}
\end{section}

\begin{section}{Finite $T$: single-magnon scattering\label{sec:T>0}}
The $S^{z}_{tot}$ symmetry allows scattering processes where no additional magnons are created. In the finite temperature case, an initial magnon of momentum ${\bf k}$ can be scattered to ${\bf k}-{\bf q}$. The one-magnon part of the scattering operator, within LSWT, takes the following form:
\begin{widetext}
    \begin{align}
        \hat{O}_{\bf q}^{(1)} = S \sum_{{\bf k} \in MBZ} &\biggl[\left( J^{AB}_{\bf 0} + J^{AB}_{\bf q} - J^{AA}_{\bf 0} - J^{AA}_{\bf q} + J^{AA}_{\bf k} + J^{AA}_{{\bf k} - {\bf q}} \right) \left( u_{\bf k} u_{{\bf k}- {\bf q}}+  v_{\bf k} v_{{\bf k} - {\bf q}} \right) \biggr. \nonumber \\
        &\biggl. -\left( J^{AB}_{\bf k}+J^{AB}_{{\bf k}-{\bf q}} \right) \left( u_{\bf k} v_{{\bf k}-{\bf q}}+v_{\bf k} u_{{\bf k}-{\bf q}} \right) \biggr] \left( \alpha^{\dag}_{{\bf k}-{\bf q}} \alpha_{\bf k} + \beta^{\dag}_{{\bf k}-{\bf q}} \beta_{\bf k} \right).
    \end{align}
\end{widetext}
We choose to concentrate on the basic case where the only non-vanishing interaction is the nearest-neighbors coupling $J$, for a 2D Heisenberg antiferromagnet with $S=1/2$.

In the low temperature regime, a few magnons of low momentum ${\bf k}$ are
present in the system. Their energy can be approximated for $T \rightarrow 0$ by letting ${\bf k} \rightarrow 0$: $\epsilon_{\bf k} \approx \sqrt{2} J |{\bf k}|$. In this limit $u_{\bf k}$ and $v_{\bf k}$ can be substituted by the following approximate expressions:
\begin{equation}
\begin{array}{l}
    u_{\bf k} \approx \frac{1}{\sqrt{\sqrt{2} |{\bf k}|}} (1+\frac{\sqrt{2}}{4} |{\bf k}|),\\
    v_{\bf k} \approx \frac{1}{\sqrt{\sqrt{2} |{\bf k}|}} (1-\frac{\sqrt{2}}{4} |{\bf k}|).
    \end{array}\label{uvapprox}
\end{equation}

In order to calculate the one-magnon contribution to the cross section, we have to evaluate the scattering amplitude expressed by Eq.~(\ref{eq:Afi}). In the low temperature case we can consider a one-magnon initial state $\ket{i}=\alpha_{{\bf k}}^{\dag} \ket{0}$.~\cite{note}  The only contribution to $A^{(1)}_{fi}$ comes from the final state with a single magnon of momentum $\bf k-\bf q$
\begin{align}
A^{(1)}_{fi}&= S \biggl[ (J_{\bf 0} + J_{\bf q}) (u_{\bf k} u_{\bf k-\bf q}+ v_{\bf k} v_{\bf k-\bf q}) \biggr. \nonumber \\
      &\biggl. \;\;\;\;-(J_{\bf k}+J_{\bf k-\bf q}) (u_{\bf k} v_{\bf k-\bf q}+  v_{\bf k} u_{\bf k-\bf q}) \biggr] \nonumber \\
      &\approx \frac{S}{\sqrt{2\sqrt{2}}} \left( J_{\bf 0} + J_{\bf q} \right)(u_{\bf q}-v_{\bf q})\sqrt{|\bf k|}
\end{align}
where we used the condition $|\bf k| \ll |\bf q|$ and inserted the expressions of Eqs.~(\ref{uvapprox}) for $u_{\bf k}$ and $v_{\bf k}$, retaining the leading order term in $|\bf k|$.

These approximations allow the analytic evaluation of the scattering intensity. At finite $T$, the cross section is given by
\begin{equation}
    \left. \frac{d^2\sigma^{(1)}}{d\Omega d\omega} \right|_{\text{res}} \propto \sum_{i,f} \frac{1}{e^{\beta E_i}-1} \left| A^{(1)}_{fi} \right|^2 \delta(\omega -E_f+ E_i).
\end{equation}
For $\textbf {k}\approx 0$, and by taking the continuum limit, we obtain
\begin{equation}
    \frac{d^2\sigma^{(1)}}{d\Omega d\omega}\propto P(\textbf q)\int_{MBZ} dk_x dk_y \frac{|\bf k|}{e^{\beta \epsilon_{\bf k}}-1} \delta(\omega -\epsilon_{\bf k-\bf q}+\epsilon_{\bf k}),
\end{equation}
where we defined $P(\textbf q )=S^2 \left( J_{\bf 0} + J_{\bf q} \right)^2 (u_{\bf q}-v_{\bf q})^2$. In the low
temperature limit, the Bose factor goes to zero rapidly for high $|{\bf k}|$, so the only substantial contribution to the integral comes from $|{\bf k}| \approx 0$. Therefore we can extend the domain of integration to the entire $k$-space. Replacing $\epsilon_{\bf k}$ with its approximate expression in the limit of low $|\bf k|$, and assuming polar coordinates, we obtain
\begin{equation}
    \frac{d^2\sigma^{(1)}}{d\Omega d\omega} \propto P(\textbf q)\int_{0}^{\infty} dk \frac{k^2}{e^{\beta \sqrt{2} J k}-1} \delta(\omega-\epsilon_{\bf q}+\sqrt{2} J k)\label{1magcross}
\end{equation}
Note that we used the replacement $\epsilon_{{\bf k} - {\bf q}} \rightarrow \epsilon_{{\bf q}}$, which breaks down at ${\bf q} = {\bf 0}$ and the BZ corners. This integral can simply evaluated to be
\begin{equation}
    \frac{d^2\sigma^{(1)}}{d\Omega d\omega} \propto P({\bf q}) \frac{\left( \omega - \epsilon_{\bf q} \right)^2}{e^{-\beta \left( \omega - \epsilon_{\bf q} \right)}-1} \theta \left( \epsilon_{\bf q} - \omega\right), \label{crosssecT}
\end{equation}
and the spectral weight for $T/J \ll 1$ is
\begin{equation}
    W_1 = \int \frac{d^2\sigma^{(1)}}{d\Omega d\omega} d\omega \propto P({\bf q}) \left(\frac{1}{\beta J}\right)^3.\label{eq:weighT}
\end{equation}
The $T^3$ behavior also shows up in the numerical evaluation of $W_1$ (without assuming $|{\bf k}| \ll |{\bf q}|$), as shown in Fig.~\ref{fig:ratio} as a function of the transferred momentum ${\bf q}$, for various temperatures (dashed lines). According to the considerations discussed in the previous section, the RIXS intensity is vanishing for $(\pi,\pi)$. The average peak position and the peak width are expected to be modified as a function of temperature. We can easily estimate these modifications by evaluating the first moment
\begin{equation}
    \langle\omega_{\max}\rangle \approx \epsilon_{\bf q} - \frac{\pi^4}{30 \zeta (3)} T,
\end{equation}
and the variance
\begin{equation}
    \langle\omega_{\max}^2\rangle - {\langle\omega_{\max}\rangle}^2 \propto T^2.
\end{equation}
We conclude that the peak position is shifted from $\epsilon_{\bf q}$ towards lower values of $\omega$, by an amount that grows linearly with $T$ and at the same time the peak broadens proportional to $T$.

We now determine the relative intensity of the one- and two-magnon scattering processes. Even if a direct comparison is
not possible, since the one-magnon and the two-magnon peaks occur at different lost energies $\omega$, it is useful to compare the one-magnon and the two-magnon total spectral weight for the $2D$ Heisenberg antiferromagnet. The latter is evaluated numerically at $T=0$, and the former at various temperatures without making the approximation ${\bf k}-{\bf q} \approx -{\bf q}$. In Fig.~\ref{fig:ratio} we plot the two-magnon (solid line) and the one-magnon weight for different temperatures (dashed lines). At room temperature, the one-magnon weight is one or two orders of magnitude smaller for almost every value of $\bf{q}$ and is expected to decrease with decreasing $T$, according to Eq. (\ref{eq:weighT}). This
allows us to conclude that the two-magnon scattering is the dominant process at low temperatures. A rough estimate for the
temperature at which the one-magnon process becomes significant gives a value of $\sim 1$ eV in the case of La$_2$CuO$_4$, which is well above room temperature. These results support the conclusion that two-magnon scattering dominates the magnetic RIXS intensities in this material observed by J.P. Hill and coworkers~\cite{Hill_tbp}. In other materials this of course needs not necessarily be so, depending on the temperature at which the experiments are performed. One can expect for instance interesting RIXS scattering signals from high temperature paramagnons.

\begin{figure}
\begin{center}
\includegraphics[angle=270,width=0.9\columnwidth]{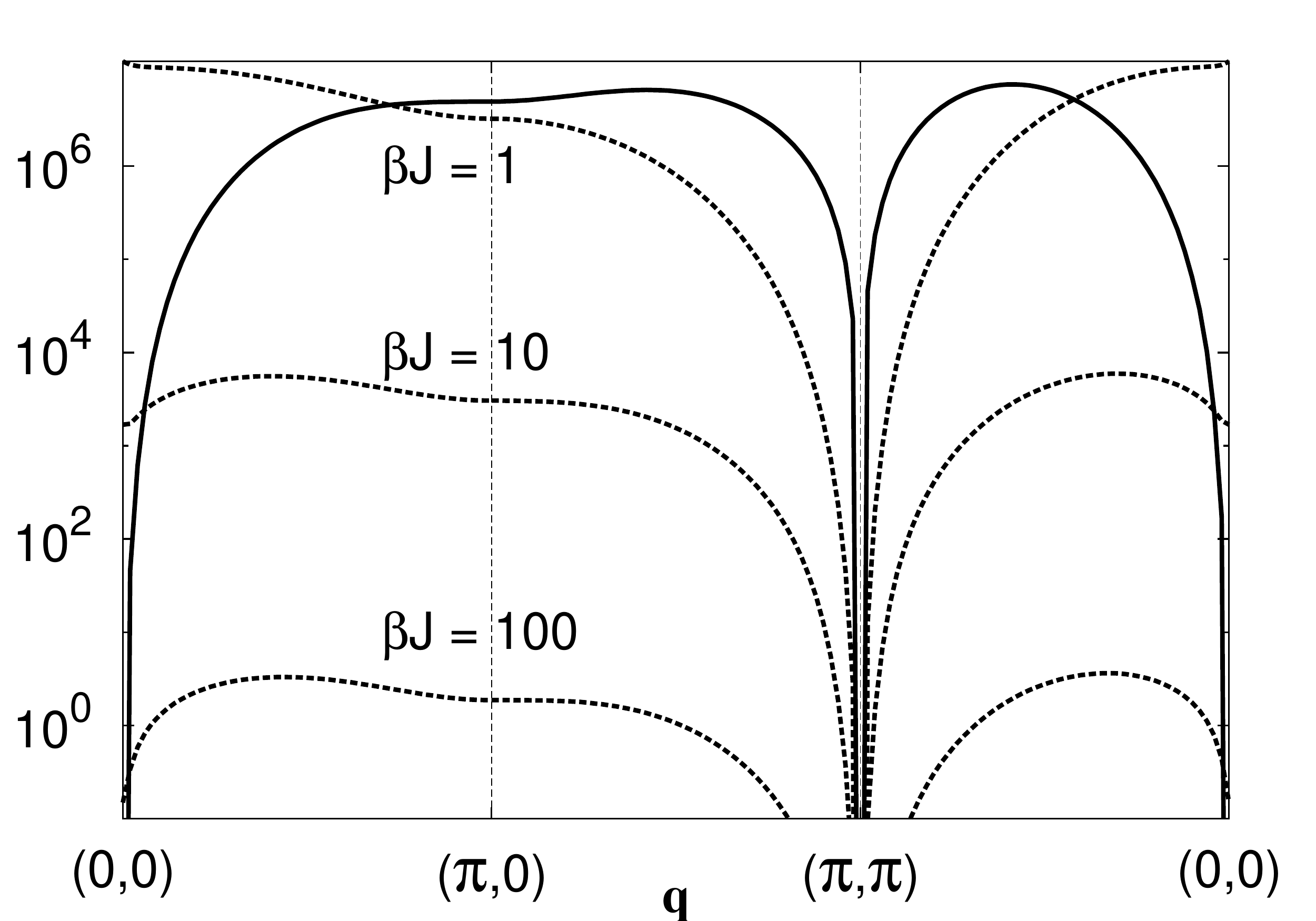}
\caption{Comparison between spectral weight for single-magnon scattering $W_1$ (dashed lines) for various  temperatures and zero temperature two-magnon scattering $W_2$ (solid line), all obtained numerically. The figure displays the $T^3$ behavior from Eq.~(\ref{eq:weighT}) for the single-magnon intensity. For La$_2$CuO$_4$  $J \approx 146$ meV, and at room temperature we have $\beta_{\text{rt}}J \approx 5.8$.\label{fig:ratio}}
\end{center}
\end{figure}
\end{section}

\begin{section}{Leading correction to ultrashort lifetime approximation}
The ultrashort core-hole lifetime (UCL) expansion offers a systematic way of calculating the Kramers-Heisenberg relation Eq.~(\ref{eq:kramersheisenberg}). In this section we calculate the leading correction term to the two-magnon cross section in the UCL approximation. This is especially relevant at ${\bf q} = (0,0)$ where the intensity is vanishing to first order, but non-zero to second order. The leading order correction is taken into account by including all terms up to $\mathcal{O}((\eta J/\Gamma)^2)$ in Eq.~(\ref{eq:Hintapprox}). Again we can include a number of extra correction terms by using an expansion of the type
\begin{align}
    \sum_{l=1}^{\infty} \frac{H_{\text{int}}^l}{\Gamma^l} \approx &\sum_{l=1}^{\infty} \left( \frac{H_0^l}{\Gamma^l} + \frac{H_0^{l-1} H'}{\Gamma^l} \right) + \sum_{l=2}^{\infty} \frac{H_0^{l-2}(H')^2}{\Gamma^l} \nonumber \\
    &+ \mathcal{O}\left( (\eta J/\Gamma)^3 \right).
\end{align}
The contribution of the last term to the UCL scattering amplitude is
\begin{equation}
    \frac{\omega_{\text{res}}}{\Gamma^2} \frac{\eta^2}{i \Gamma + \omega} \bra{f}\sum_i e^{i {\bf q}\cdot {\bf R}_i}\sum_{j,k}J_{ij}J_{ik}({\bf S}_i \cdot {\bf S}_j)({\bf S}_i \cdot {\bf S}_k)\ket{i} \label{eq:Aficorr}
\end{equation}
This scattering amplitude that corresponds to this term is non-zero at ${\bf q} = {\bf 0}$, which can be easily checked in linear spin-wave theory. The reason is that the resulting scattering operator at zero transferred momentum does not commute with the Hamiltonian. For the LSW analysis we make use of the identity
\begin{align}
    \sum_{j,k} J_{ij}J_{ik} &({\bf S}_i \cdot {\bf S}_j)({\bf S}_i \cdot {\bf S}_k) = \frac{1}{4} \sum_{j \neq k} J_{ij}J_{ik}{\bf S}_j \cdot {\bf S}_k \nonumber \\
    &- \frac{1}{2} \sum_j J_{ij}^2 {\bf S}_i \cdot {\bf S}_j + \text{const}.
\label{eq:fourspindecompose}
\end{align}
We drop the constant because it does not contribute to inelastic scattering. For simplicity, we only take nearest neighbor interactions into
account. The last term in Eq.~ (\ref{eq:fourspindecompose}) is
proportional to the first order result for the scattering
amplitude, which has already been analyzed in LSWT. The other term
can be treated in LSWT too, and yields a two-magnon contribution
to the scattering amplitude at zero temperature of:
\begin{align}
 - \frac{\omega_{\text{res}}}{4 \Gamma^2} &\frac{\eta^2 J^2}{i\Gamma + \omega} \sum_{\bf k} \bra{f} f({\bf k},{\bf q})
 \times \nonumber \\
 &(u_{\bf k}v_{{\bf k}+{\bf q}}+u_{{\bf k}+{\bf q}}v_{\bf k}) \alpha^{\dag}_{\bf k} \beta^{\dag}_{-{\bf k}-{\bf q}} \ket{0}\label{eq:secondorder}
\end{align}
with $f({\bf k},{\bf q}) = -6(\cos q_x + \cos q_y) + 4 \cos k_x \cos (k_y + q_y) + 4 \cos k_y \cos (k_x + q_x) + 2 \cos (2k_x+q_x) + 2 \cos (2 k_y + q_y)$. Since the phase of the first order amplitude differs from the second order amplitude by $\pi/2$, there is no interference of these terms. The consequence is that the leading corrections to the first order scattering intensity are down by a factor $(\eta J/\Gamma)^2 \approx 0.06$ for the well-screened intermediate state. This makes the ultrashort core-hole lifetime approximation a viable way of computing magnetic RIXS spectra. The contribution Eq.~(\ref{eq:Aficorr}) is shown in Fig.~\ref{fig:secondorder}(a), and the full cross section in Fig.~\ref{fig:secondorder}(b). Only at ${\bf q}={\bf 0}$ there is an appreciable difference from the first order result shown in Fig.~\ref{fig:rixsdos} (a). At ${\bf q} = (\pi,\pi)$, there is again no intensity, which can be understood by the same argument as for the first order result in section \ref{sec:T=0}.

\begin{figure}
\begin{center}
\begin{tabular}{c}
\begin{minipage}{\columnwidth}
\hspace{-1cm}\includegraphics*[width=\columnwidth]{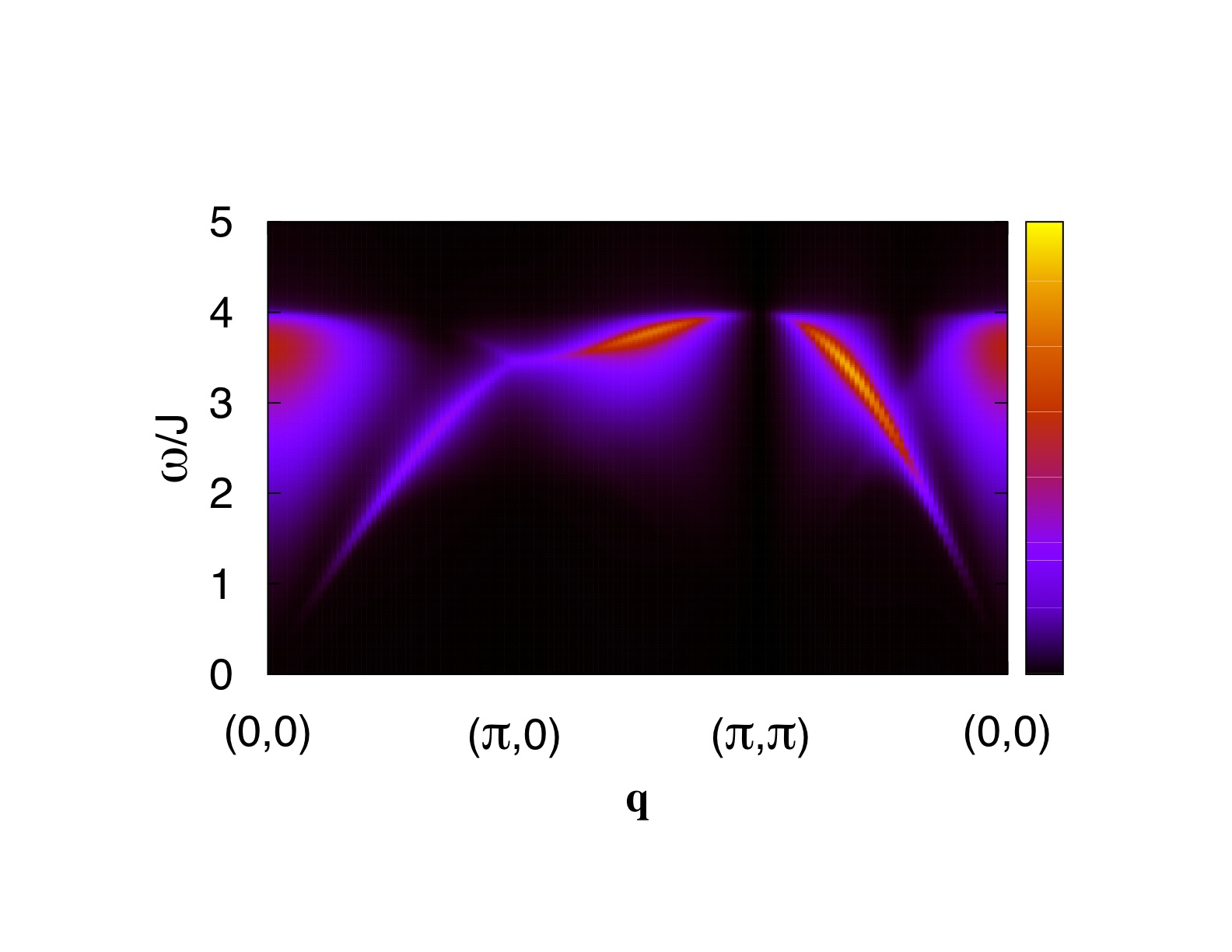}\vspace{-1cm}
\begin{center}(a)\end{center}
\end{minipage}
\\
\begin{minipage}{\columnwidth}
\hspace{-1cm}\includegraphics*[width=\columnwidth]{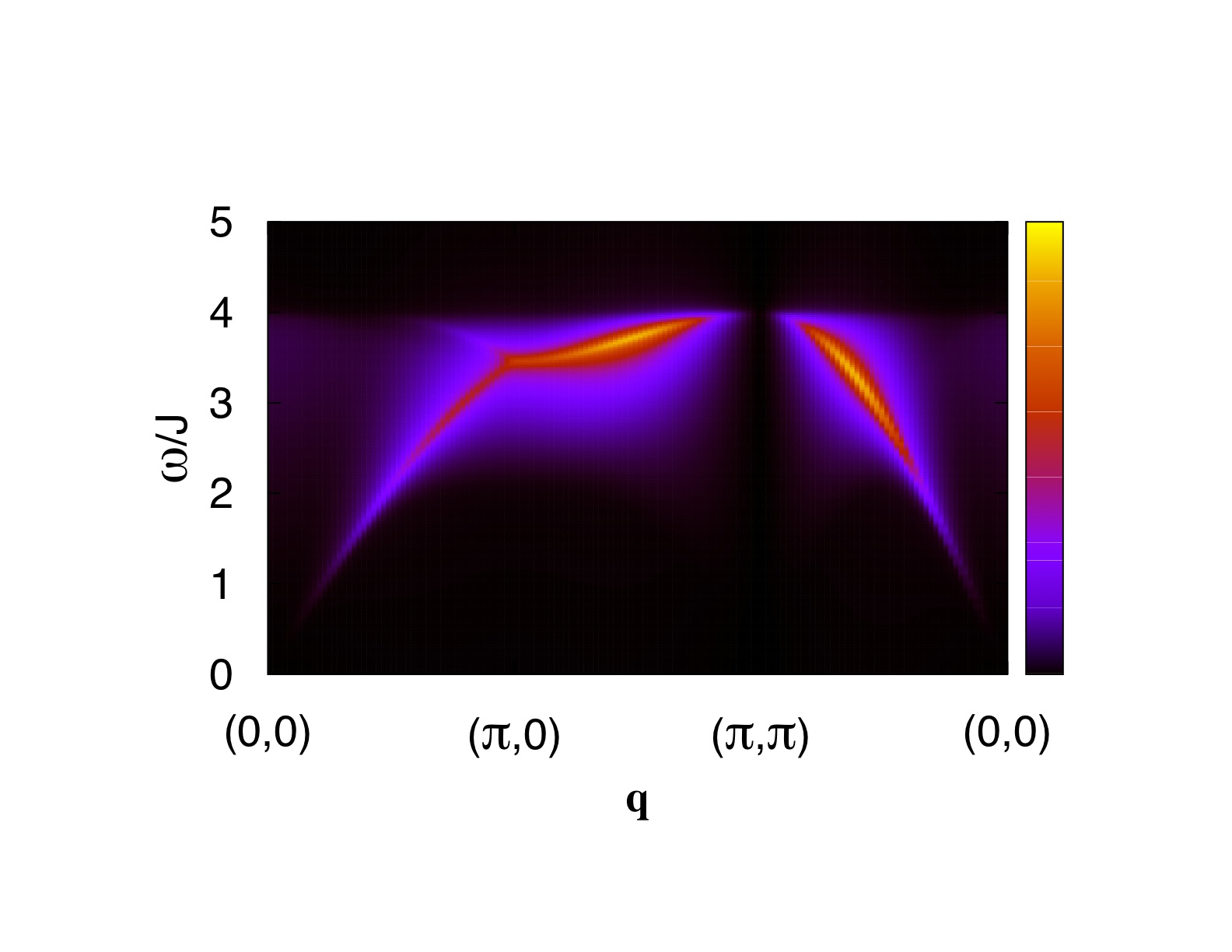}\vspace{-1cm}
\begin{center}(b)\end{center}
\end{minipage}
\end{tabular}
\end{center}
\caption{The leading order correction to the scattering amplitude does not interfere with the first order. Fig. (a) shows the contribution to the cross section from Eq.~(\ref{eq:Aficorr}). The full, corrected cross section is shown in Fig.~(b). There is an appreciable correction only at ${\bf q} = {\bf 0}$.\label{fig:secondorder}}
\end{figure}
\end{section}

\begin{section}{Conclusions\label{sec:conc}}
We derived the two-magnon scattering cross section which is measured in magnetic RIXS at the Cu $K$-edge, taking advantage of a series expansion in the ultrashort core-hole lifetime (UCL) of the intermediate state. In the context of LSWT, we calculated the magnetic RIXS spectrum for a 2D $S=1/2$ Heisenberg antiferromagnet, in the more general case where the superexchange is not limited to nearest neighbors. Our results strongly suggest a multi-magnon scattering scenario, where two-magnon excitations are created in the system as a consequence of the modifications in the superexchange interaction induced by the core-hole potential.

Our results for the two-magnon scattering agree very well with experimental data on La$_2$CuO$_4$. The vanishing of the RIXS intensity for the elastic case ${\bf q} = (0,0)$ and the antiferromagnetic point $\textbf{q}=(\pi,\pi)$ is recovered. The latter feature turns out to be a consequence of an underlying symmetry property of the scattering operator and does not depend on the range of the exchange interaction. The excellent quantitative agreement between our results and experiments is testified by the occurrence of an intense peak at $\textbf{q}=(\pi,0)$ for $\omega \approx 500$ meV. We have generalized the theory to include also finite-temperature scattering, for which we find that also one-magnon processes contribute.  For La$_2$CuO$_4$ at room temperature the single magnon spectral weight is very small compared to two-magnon scattering.

The subleading order in the UCL expansion of the cross section is shown to be of order $\mathcal{O}((\eta J/\Gamma)^2)$ smaller than the first order result. This makes the UCL approximation a rigorous method for this case to calculate the Kramers-Heisenberg relation. The introduction of longer range interactions (according to data from neutron experiments) improves the correspondence between theory and magnetic RIXS experiments on La$_2$CuO$_4$. The generalization of the analysis to doped systems will be an interesting next step towards understanding multi-spin correlations in the spin liquid phase of the high-T$_c$ superconductors.
\end{section}

\begin{section}{Acknowledgments}
We thank Michel van Veenendaal for stimulating discussions and John P. Hill for also sharing unpublished data with us.
We gratefully acknowledge support from the Argonne National Laboratory Theory Institute, Brookhaven National Laboratory (DE-AC02-98CH10996) and the Dutch Science Foundation FOM. This paper was supported in part by the National Science Foundation under Grant No. PHY05-51164.

\end{section}

\end{document}